# Formation of the binary pulsars J1141–6545 and B2023+46

Melvyn B. Davies[1], Hans Ritter[2], and Andrew King[1]

[1] *Department Physics & Astronomy, University of Leicester, Leicester LE1 7RH, UK*
[2] *Max–Planck–Institut für Astrophysik, Karl–Schwarzschild–Str. 1, D–85740 Garching, Germany*

26 October 2018

**ABSTRACT**
The binaries PSR J1141–6545 and PSR B2303+46 each appear to contain a white dwarf which formed before the neutron star. We describe an evolutionary pathway to produce these two systems. In this scenario, the primary transfers its envelope onto the secondary which is then the more massive of the two stars, and indeed sufficiently massive later to produce a neutron star via a supernova. The core of the primary produces a massive white dwarf which enters into a common envelope with the core of the secondary when the latter evolves off the main sequence. During the common envelope phase, the white dwarf and the core of the secondary spiral together as the envelope is ejected. The evolutionary history of PSR J1141–6545 and PSR B2303+46 differ after this phase. In the case of PSR J1141–6545, the secondary (now a helium star) evolves into contact transferring its envelope onto the white dwarf. We propose that the vast majority of this material is in fact ejected from the system. The remains of the secondary then explode as a supernova producing a neutron star. Generally the white dwarf and neutron star will remain bound in tight, often eccentric, systems resembling PSR J1141–6545. These systems will spiral in and merge on a relatively short timescale and may make a significant contribution to the population of gamma ray burst progenitors. In PSR B2303+46, the helium-star secondary and white dwarf never come into contact. Rather the helium star loses its envelope via a wind, which increases the binary separation slightly. Only a small fraction of such systems will remain bound when the neutron star is formed (as the systems are wider). Those systems which are broken up will produce a population of high–velocity white dwarfs and neutron stars.

## 1 INTRODUCTION

Observations suggest that in the two binaries PSRs B2303+46 and J1141–6545, the companions of the observed pulsars are white dwarfs which appear to have been formed *before* the neutron stars (van Kerkwijk & Kulkarni 1999; Kaspi et al 2000; Manchester et al 2000).

If the white dwarf were the remains of a former donor star of the neutron star, we would expect the white dwarf to be in a circular orbit. However B2023+46 has an eccentricity of 0.658, so we must conclude that the white dwarf was made first and the current observed eccentricity is derived from the kick received when the neutron star was formed.

J1141–6545 is a non-recycled pulsar with a massive companion in an eccentric orbit (Kaspi et al 2000; Manchester et al 2000). If the pulsar had formed first, we would expect it to have been spun up (ie recycled) when the secondary evolved into contact and transferred material onto it, ultimately leaving either a second neutron star or a white dwarf. If the companion produced a white dwarf after the birth of the neutron star, we would also expect the orbit today to be circular, whereas it has a measured eccentricity of 0.172. Hence it seems likely that in J1141–6545 the companion is also probably a massive white dwarf which was formed before the neutron star. It should be noted that although many of the basic properties of J1141–6545 and B2023+46 are similar, their orbital periods differ by a factor of 60. Accounting for this fact is one of the problems we are going to address in this paper.

We consider an evolutionary scenario for the production of such white-dwarf neutron-star (WD–NS) binaries. In this scenario, the original primary transfers its envelope to the secondary conservatively leaving only the helium-star core. This helium star then evolves, filling its Roche lobe, leading to a second phase of mass transfer where the en-velope of the helium star is transferred to the secondary (which is still on the main sequence). In the process the primary becomes a white dwarf. In sufficiently wide binaries, when the secondary evolves into contact, the ensuing phase of mass transfer will produce a common-envelope system in which the white dwarf and the helium-star core of the secondary will spiral together as the common envelope of gas is ejected from the system. The post-common-envelope system will consist of the white dwarf and the helium star in a tight binary. Providing sufficient mass has been transferred to the secondary during earlier phases in the binary evolution, the helium star may be massive enough to produce a neutron star via a supernova explosion.

This scenario has been considered elsewhere (see for example Portegies–Zwart & Yungelson 1999; Brown et al 2000; Tauris & Sennels 2000). Here we extend the earlier work by considering the role of the helium star radius in the evolution of the system after the common envelope phase which will depend on the separation of the binary and the maximum size of the pre-supernova helium star.

Systems similar to B2023+46 will be produced in relatively wide systems, where the helium star avoids filling its Roche lobe as it expands. Instead it loses mass via a wind. For closer binaries (ie those producing systems similar to J1141–6545), the helium star in the post-common-envelope system will fill its Roche lobe. In this paper, we propose that a phase of mass transfer follows where the helium star envelope is transferred to the white dwarf at highly super-Eddington rates. Rather than being accreted, we propose that this material is ejected from the system.

In Section 2, we begin by describing a simplified version of the scenario described above where we neglect the radius of the final helium star. In Section 3 we compute the population of binaries produced using this simplified scenario. In Section 4 we consider the role of the helium-star radius on the binary evolution. In Section 5 we consider the inferred birthrate of B2023+46–like and J1141–6545–like systems. We also consider the timescale for the merger of the tighter systems and the importance of the kick velocities received by the binaries on formation of the neutron star. We summarise our results in Section 6.

## 2 EVOLUTIONARY SCENARIO

The key feature of both J1141–6545 and B2303+46 is that the neutron star is *younger* than the white dwarf. This implies that the star which produced the neutron star in each system must have originally been the secondary. Yet the primary was only massive enough to produce a white dwarf. Mass transfer from the primary to the secondary must have increased its mass sufficiently to yield a supernova. These requirements greatly restrict the binary parameter space which may produce such neutron star–white dwarf (WD–NS) systems. The evolutionary sequence envisaged to produce WD–NS binaries is outlined below:

We consider a binary of initial separation, $a_i$, and total mass contained in the system $\mathcal{M}_i = M_{1,i} + M_{2,i}$.

The first phase of mass transfer occurs when the primary fills its Roche lobe just off the main sequence. During this phase of conservative, radiative mass transfer (early Case B), the primary transfers its entire envelope to the secondary. Only the helium core of the primary remains. The masses of the two stars, $M_{1,B}$ and $M_{2,B}$, after this phase of mass transfer are given by

$$M_{1,B} = M_{He}(M_{1,i}) = a M_{1,i}^b \quad (1)$$

$$M_{2,B} = \mathcal{M}_i - M_{1,B} \quad (2)$$

where $a = 0.125$ and $b = 1.4$ (van den Heuvel 1994). The ratio of the separation after this phase to the initial separation is given by

$$F_{i \to B} = \frac{a_B}{a_i} = \frac{M_{1,i}^2 M_{2,i}^2}{M_{1,B}^2 M_{2,B}^2} \quad (3)$$

This helium star then evolves filling its Roche lobe leading to the second phase of mass transfer (Case BB) where the primary transfers the envelope of the helium star to the secondary (which is still on the main sequence). In a calculation of Case BB evolution by Delgado & Thomas (1981), a 2 $M_\odot$ helium star initiates mass transfer when its radius is about 20 $R_\odot$ and terminates mass transfer at a radius of about 50 $R_\odot$ yet it is stripped down to the CO core. We therefore assume here that after this phase of mass transfer the primary becomes a CO or ONeMg white dwarf. The masses of the two stars, $M_{1,BB}$ and $M_{2,BB}$, after this phase of mass transfer are given by

$$M_{1,BB} = M_{CO}(M_{1,B}) = M_{WD} \quad (4)$$

$$M_{2,BB} = \mathcal{M}_i - M_{1,BB} \quad (5)$$

The ratio of the separation after the Case BB mass transfer to the initial separation is given by

$$F_{i \to BB} = \frac{a_{BB}}{a_i} = \frac{M_{1,i}^2 M_{2,i}^2}{M_{1,BB}^2 M_{2,BB}^2} \quad (6)$$

The secondary will evolve into contact on its nuclear evolutionary timescale. If the system is sufficiently wide, this third phase of mass transfer will be convective Case B which will lead to a common–envelope phase during which the white dwarf and the helium–star core of the secondary spiral together as the common envelope of gas is ejected from the system. Alternatively, the a common–envelope phase will be initiated from radiative, Case B mass transfer, followed by a delayed dynamical instability (DDI) if the mass ratio is large enough (Hjellming 1989, Kalogera & Webbink 1996). The masses of the two stars after this phase, $M_{1,CE}$ and $M_{2,CE}$, are given by

$$M_{1,CE} = M_{1,BB} = M_{WD} \quad (7)$$

$$M_{2,CE} = M_{He}(M_{2,BB}) = a M_{2,BB}^b \quad (8)$$

The inspiral during the common envelope phase may be computed by equating the change in orbital energy of the two stars to the binding energy of the envelope up to an efficiency $\alpha_{CE}$, ie $E_{env} = \alpha_{CE} \Delta E$. Here $E_{env}$ can be written in the following form (Webbink 1984)

$$E_{env} = \frac{G M_{2,BB}(M_{2,BB} - M_{2,CE})}{\lambda_{CE} f_2(q_{BB}) a_{BB}} \quad (9)$$

where the radius of star 2 when it fills its Roche lobe is given by $R_{2,BB} = f_2(q_{BB}) a_{BB}$, where $q_{BB} = M_{1,BB}/M_{2,BB}$ and $f_2(q) = 0.49 q^{2/3}/(0.6 q^{2/3} + \ln(1 + q^{1/3}))$ (Eggleton 1983).

Combining equation (9) with the expression for $\Delta E$, after some rearrangement we arrive at the following expression for the inspiral during the common envelope phase

$$F_{\rm BB \to CE} = \frac{a_{\rm CE}}{a_{\rm BB}} = \left(\frac{2M_{2,\rm BB}(M_{2,\rm BB} - M_{2,\rm CE})}{\alpha_{\rm CE}\lambda_{\rm CE}f_2(q_{\rm BB})M_{2,\rm CE}M_{1,\rm BB}} + \frac{M_{2,\rm BB}}{M_{2,\rm CE}}\right)^{-1} \quad (10)$$

We may combine equations (6) and (10) to produce an expression relating the initial separation of the binary to that after the common envelope phase.

$$F_{\rm i \to CE} = \frac{a_{\rm CE}}{a_{\rm i}} = F_{\rm i \to BB} \cdot F_{\rm BB \to CE} \quad (11)$$

The post-common-envelope system will consist of the white dwarf and the helium star in a tight binary. If the helium star is sufficiently massive, it will explode as a supernova, producing a neutron star. The instantaneous mass-loss together with the kick the neutron receives at birth will unbind some of these binaries, while others will remain bound. By sampling the neutron star kick distribution of Hansen & Phinney (1997), and allowing for the effects of mass-loss during the supernova, we are able to produce a population of WD–NS binaries and plot their binary properties and system kick velocities, given a population of pre-supernova binaries.

Before we inspect the population of WD–NS systems produced in the scenario outlined above, we must consider the constraints on the properties of the original pre-evolution binaries.

### 2.1 Constraints on initial masses

We require that the mass of the primary should be insufficient to produce a supernova on its own. In other words the mass of the compact object produced after the case BB mass transfer (as given in equation [4]) does not exceed the Chandrasekhar mass:

$$M_{\rm WD} = M_{\rm CO}(aM_{1,\rm i}^b) < M_{\rm CH} \quad (12)$$

By definition the primary mass exceeds that of the secondary, ie

$$M_{1,\rm i} > M_{2,\rm i} \quad (13)$$

In order to avoid the onset of a delayed dynamical instability when mass is transferred from the primary to the secondary we require

$$M_{2,\rm i} \gtrsim M_{1,\rm i}/3 \quad (14)$$

In addition, we require that the secondary does ultimately produce a supernova. In other words, $M_{\rm CO}(M_{2,\rm CE}) > M_{\rm CH}$. As $M_{2,\rm CE} = aM_{2,\rm BB}^b = a(M_{1,\rm i} + M_{2,\rm i} - M_{\rm WD})^b$, rearrangement yields

$$M_{2,\rm i} > \left(\frac{M_{\rm He}(M_{\rm CH})}{a}\right)^{1/b} + M_{\rm WD} - M_{1,\rm i} \quad (15)$$

The constraints on the initial masses given in equations (12) – (15) are illustrated in Fig. 1. The maximum allowed mass for the primary is set by equation (12) and is given by $M_{1,\rm i} = 8.3 M_\odot$. The upper mass limit for $M_{2,\rm i}$ is given by equation (13), while the lower-mass limit is given by equation (15) for lower values of $M_{1,\rm i}$ and by equation (14) for larger values of $M_{1,\rm i}$.

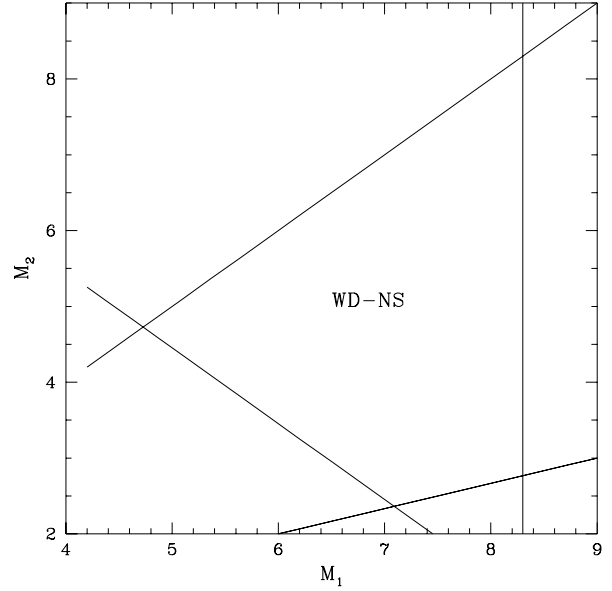

**Figure 1.** Plot of initial primary mass $M_1$ as a function of secondary mass $M_2$ (both in solar units) showing constraints for the formation of a WD–NS binary as described in section 3.

### 2.2 Constraints on initial separations

The evolutionary pathway described above places a number of constraints on the initial separation, $a_{\rm i}$, of the binary if a WD–NS system is to be produced.

For the first phase of mass transfer to be radiative Case B mass transfer, we require that the radius of the primary when it first fills its Roche lobe, $R_{1,\rm i}$, is larger than the minimum radius required for Case B transfer, $R_{\rm min,B}(M_{1,\rm i})$, and smaller than the maximum radius allowed for radiative case B mass transfer, $R_{\rm max,rB}(M_{1,\rm i})$. In other words

$$\frac{R_{\rm min,B}(M_{1,\rm i})}{f_1(q_{\rm i})} < a_{\rm i} < \frac{R_{\rm max,rB}(M_{1,\rm i})}{f_1(q_{\rm i})} \quad (16)$$

where $R_{1,\rm i} = f_1(q_{\rm i})a_{\rm i}$ with $q_{\rm i} = M_{1,\rm i}/M_{2,\rm i}$ and $f_1(q) = 0.49q^{-2/3}/(0.6q^{-2/3} + \ln(1 + q^{-1/3}))$. In order for Case BB mass transfer from the primary to occur, we require the helium star to fill its Roche lobe. In other words

$$a_{\rm i} < \frac{R_{\rm max,He}(M_{1,b})}{F_{\rm i \to B}f_1(q_{\rm B})} \quad (17)$$

where $q_{\rm B} = M_{1,\rm B}/M_{2,\rm B}$. In order for the late (convective) Case B mass transfer from the secondary to occur, leading to the onset of a common envelope phase, we require that the secondary fill its Roche lobe, but not until it has a radius larger than the maximum allowed for early (radiative) Case B mass transfer, ie

$$\frac{R_{\rm max,rB}(M_{2,\rm BB})}{F_{\rm i \to BB}f_2(q_{\rm BB})} < a_{\rm i} < \frac{R_{\rm max,B}(M_{2,\rm BB})}{F_{\rm i \to BB}f_2(q_{\rm BB})} \quad (18)$$

where $q_{\rm BB} = M_{1,\rm BB}/M_{2,\rm BB}$ and $R_{\rm max,B}(M_{2,\rm BB})$ is the maximum radius allowed for Case B mass transfer for a star of mass $M_{2,\mathrm{BB}}$. A CE envelope can also follow from a donor with a radiative envelope if the mass ratio is big enough for a delayed–dynamical instability to occur. The systems of interest here have a white–dwarf primary mass of 1–1.4 $M_\odot$ and a donor star whose mass is above the lower limit for evolving to a supernova, ie $M_2 > 8.3$ $M_\odot$. This means that the lower limit implied by equation (18) will not be a restriction to the onset of a CE phase and thus on the formation of WD–NS binaries.

We require that the helium main-sequence-star secondary does not fill its Roche lobe at the end of the common envelope phase.

$$a_\mathrm{i} > \frac{R_{\mathrm{HeMS}}(M_{2,\mathrm{CE}})}{F_{\mathrm{i}\to\mathrm{CE}} f_2(q_{\mathrm{CE}})} \qquad (19)$$

where $R_{\mathrm{HeMS}}(M_{2,\mathrm{CE}})$ is the radius of the helium main-sequence-star secondary and $q_{\mathrm{CE}} = M_{1,\mathrm{CE}}/M_{2,\mathrm{CE}}$.

## 3  EVOLUTION IGNORING THE HE STAR RADIUS

We now consider the evolution of binaries following the scenario descibed above, applying the constraints given in equations (12) – (19). The evolution of the binary separations and stellar masses were computed using equations (1) – (11). The functions $R_{\mathrm{min,B}}(M)$, $R_{\mathrm{max,rB}}(M)$, and $R_{\mathrm{max,B}}(M)$ were tabulated from Bressan et al. (1993), $R_{\mathrm{max,He}}(M)$ from Paczyński (1971) and Habets (1986), and $M_{\mathrm{CO}}(M)$ from Habets (1986). Linear interpolation was used to evaluate them for a particular mass, $M$. We note that the only information which we draw from Paczyński (1971) is $R_{\mathrm{max,He}}(M)$, and this only to compare with results of Habets (1986), on which we mostly rely. The radius of a zero-age helium main-sequence star, $R_{\mathrm{HeMS}}(M)$, can be approximated as follows (Habets 1986)

$$\log \frac{R_{\mathrm{HeMS}}}{R_\odot} \simeq -0.68 + 0.67 \log \frac{M_{\mathrm{He}}}{M_\odot} \qquad (20)$$

For illustration we consider here systems having an initial primary mass, $M_{1,\mathrm{i}} = 7.25 M_\odot$. We follow the evolution of binaries having a range of initial separations, $a_\mathrm{i}$, and secondary masses, $M_{2,\mathrm{i}}$ (within the range allowed by equations [13] – [15]). We thus synthesize a population of binaries containing a white dwarf and a helium star on the verge of exploding as a supernova. We then apply a random kick to the neutron star formed in the supernova explosion (drawn from the kick distribution of Hansen & Phinney [1997]) and combine it with the dynamical consequences of the instantaneous ejection of the helium star envelope. A fraction of the binaries will be broken up by the supernova; those that remain form our population of WD–NS systems, and should contain systems resembling both J1141–6554 and B2023+46.

The initial parameter space (ie $a_\mathrm{i}$ and $M_{2,\mathrm{i}}$) which produces WD–NS systems is shown as a dotted region in Fig. 2, where the five lines drawn are the constraints given in equations (16) – (19). Here we have assumed $\alpha_{\mathrm{CE}}\lambda_{\mathrm{CE}} = 0.5$. The range of values of $M_2$ is constrained by equations (13) and (14). The top line in Fig. 2 is the upper limit given in equation (17) (labelled $U_{17}$) and the bottom line is the the lower limit from equation (16) (labelled $L_{16}$) with other lines being labelled accordingly. We therefore see from the figure that the lower limit on initial separations is constrained by the lower limits of equations (16) and (19), ie the requirements that the first phase of mass transfer is Case B, and that the helium main–sequence star secondary does not fill its Roche lobe at the end of the common envelope phase. The upper limit on initial separations is given by the upper limit from equation (16) for $M_2 \leq 4$ M$_\odot$, ie by requiring the first phase of mass transfer from the primary to occur while the star is still radiative. For more massive secondaries, the upper limit on initial separations is given by the upper limit from equation (18), by requiring that the secondary fill its Roche lobe causing the onset of a common envelope phase. The lower limit from equation (18) divides systems entering a common envelope phase via convective mass transfer (above $L_{18}$) and via a delayed dynamical instability (below $L_{18}$).

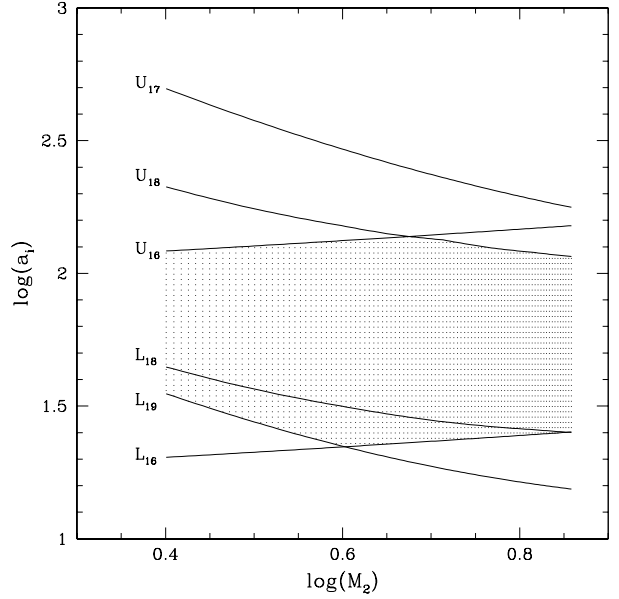

**Figure 2.** Plot of initial separation $a_i$ as a function of secondary mass $M_2$ (both in solar units) showing constraints for the formation of a WD–NS binary as described in section 3.

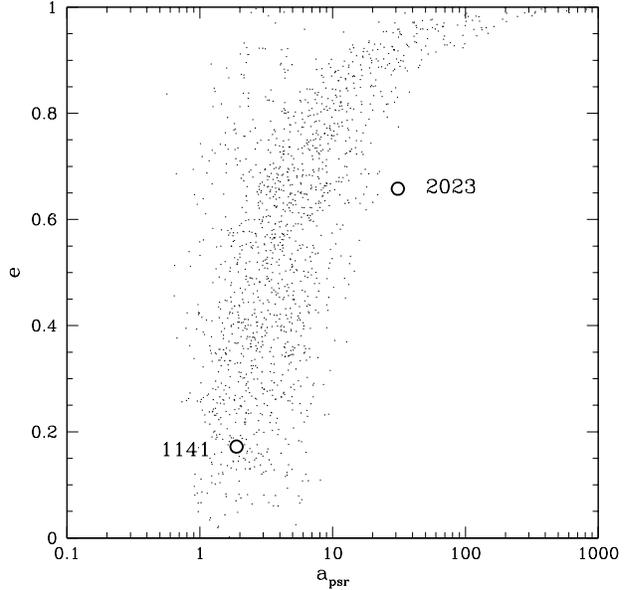

**Figure 3.** Plot of eccentricity $e$ as a function of separation, $a_{\mathrm{psr}}$ (in solar radii) for WD–NS systems produced through the evolutionary scenario described in section 3. The two open circles are the observed systems J1141–6545 and B2023+46.

The resulting population of WD–NS systems is shown in Fig. 3, where we plot the semi-major axes of the binaries, $a_{\rm psr}$, as a function of their eccentricities, $e$. The observed locations of J1141-6545 and B2303+46 are also shown as open circles (separations of 1.89 R$_\odot$ and 31.0 R$_\odot$ respectively, where we have assumed total masses of 2.3 M$_\odot$ and 2.64 M$_\odot$ respectively). We see from this plot that under the assumptions of section 2, we can easily produce J1141–6545, but that B2023+46 is wider (by a factor of about 2) than any system of a similar eccentricity produced by our population synthesis code. Wider systems will be produced if we increase the value of $\alpha_{\rm CE}\lambda_{\rm CE}$ to values in excess of unity, but then this makes it much harder to produce J1141–6545. In fact, in this approach, one cannot have both systems explained with the same value of $\alpha_{\rm CE}\lambda_{\rm CE}$. Whether $\alpha_{\rm CE}\lambda_{\rm CE} \gtrsim 1$ implies $\alpha_{\rm CE} > 1$ depends on the value of $\lambda_{\rm CE}$, which is uncertain, although values greater than unity seem to be possible (see Dewi & Tauris 2000, and Tauris & Dewi 2001).

## 4 ROLE OF THE HELIUM STAR RADIUS

### 4.1 B2023+46–like evolution

We consider now the evolution of the helium main-sequence star after the common envelope phase. In the previous section we assumed that its mass and radius did not evolve prior to the supernova explosion. Neither assumption is correct. In isolation, the helium star will evolve to a radius $R_{\rm max,He}(M_{2,\rm CE})$. If this star is to avoid filling its Roche lobe and transferring material to the white dwarf, we require

$$a_{\rm i} > \frac{R_{\rm max,He}(M_{2,\rm CE})}{F_{\rm i \to CE} f_2(q_{\rm CE})} \quad (21)$$

where $q_{\rm CE} = M_{1,\rm CE}/M_{2,\rm CE}$. The initial parameter space (ie $a_{\rm i}$ and $M_{2,\rm i}$) which satisfies equations (16) – (19) and equation (21) is shown in the dotted region in Fig. 4, where we have assumed $\alpha_{\rm CE}\lambda_{\rm CE} = 0.5$. Note that the initial phase space is much reduced compared to Fig. 2. If equation (21) is satisfied, the helium star will lose its envelope as a wind before the remaining core explodes as a supernova, where the pre-supernova core mass is given by

$$M_{2,\rm PSN} = M_{\rm CO}(M_{2,\rm CE}) \quad (22)$$

This mass loss will also affect the separation of the two stars. The relative change in separation will be given by

$$F_{\rm CE \to PSN} = \frac{a_{\rm PSN}}{a_{\rm CE}} = \frac{M_{\rm WD} + M_{2,\rm CE}}{M_{\rm WD} + M_{2,\rm PSN}} \quad (23)$$

Hence the mass loss will *increase* the separation of the two stars prior to the supernova. Continuing the evolution of the binaries as before we produce a population of post-supernova systems, as illustrated in Fig. 5. Note from this figure that B2023+46-like are on the edge (but within) the phase space of allowed systems. No systems similar to J1141–6545 are produced. It is in fact quite impossible to produce a system resembling J1141–6545 without the helium star filling its Roche lobe in the post-common-envelope phase. The timescale for a wider, more eccentric system to evolve into a J1141–6545-like system via the emission of gravitational waves is prohibitively long. We therefore conclude that this type of evolution is able to explain the production of B2023+46 but is unable to explain the origin of J1141–6545.

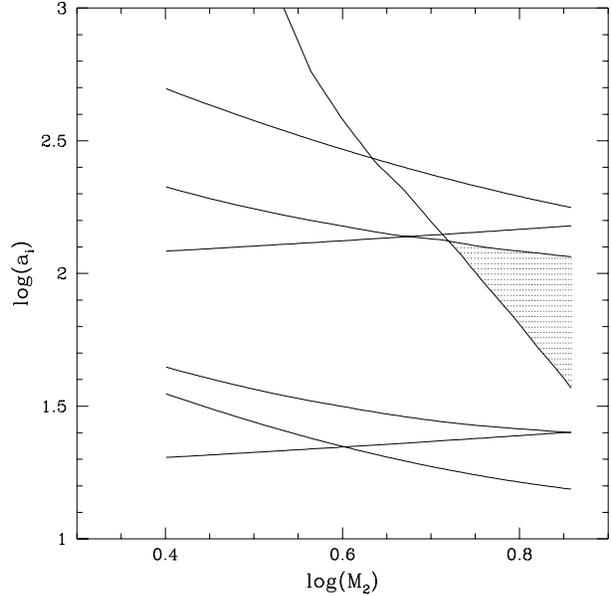

**Figure 4.** Plot of initial separation $a_i$ as a function of secondary mass $M_2$ (both in solar units) showing constraints for the formation of a WD–NS binary as described in section 4, where the actual radius of the pre-supernova helium star is considered.

### 4.2 J1141–6545–like evolution

We consider next the case where the helium star fills its Roche lobe prior to the supernova, ie

$$a_{\rm i} < \frac{R_{\rm max,He}(M_{2,\rm CE})}{F_{\rm i \to CE} f_2(q_{\rm CE})} \quad (24)$$

where $q_{\rm CE} = M_{1,\rm CE}/M_{2,\rm CE}$. A number of outcomes are then possible. If a second common envelope phase were to ensue, the core of the helium star and the white dwarf would probably merge and we would be left with a single object rather than a binary. Alternatively, one might imagine a phase of mass transfer similar to that envisaged for Cygnus X-2 (King & Ritter 1999, Kolb et al. 2000) where material was transferred rapidly from a donor on to a compact object (in that case, a neutron star) but rather than being accreted, the vast bulk of the material was ejected from the system. Taking the data from Habets (1986), we estimate that $\dot{M}_{\rm KH} = M/\tau_{\rm KH}$ for He-stars in the mass range of interest $\sim 2.5$M$_\odot - \sim 4$M$_\odot$ is of the order of $\sim 10^{-3}$M$_\odot$yr$^{-1}$ if the radius of the He-star is $R_{\rm He} \approx 3R_\odot$. At the same time the

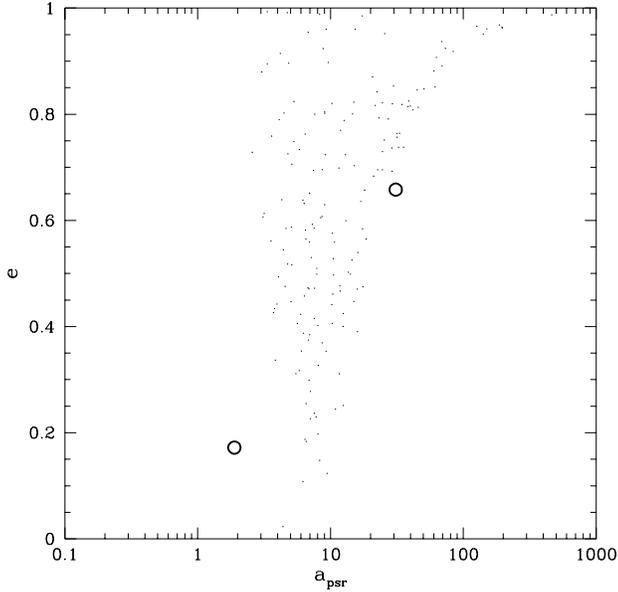

**Figure 5.** Plot of eccentricity $e$ as a function of separation, $a_{\rm psr}$ (in solar radii) for systems produced without mass transfer after the common-envelope phase as described in section 4. As in Figure 3, the two open circles are the observed systems J1141–6545 and B2023+46.

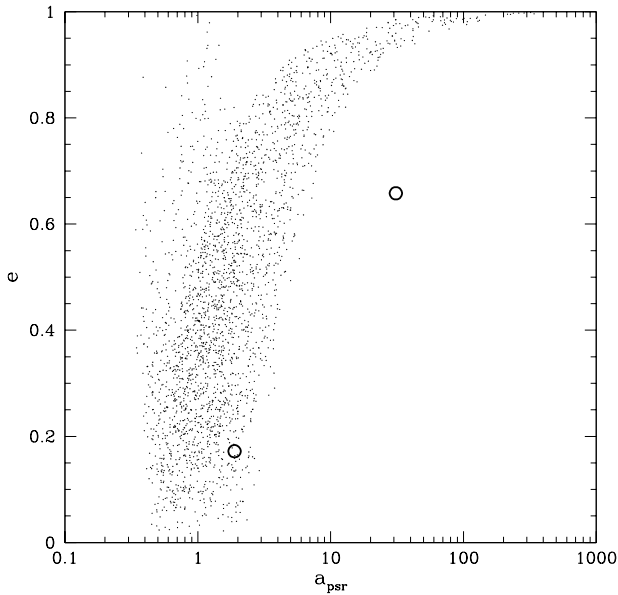

**Figure 6.** Plot of eccentricity $e$ as a function of separation (in solar radii) for WD–NS systems undergoing mass transfer after the common envelope phase as described in section 4. As in Figure 3, the two open circles are the observed systems J1141–6545 and B2023+46.

Eddington accretion rate for a WD of mass $M_{\rm WD} = 1.2 M_\odot$ is $\dot{M}_{\rm Edd} \approx 6.4 \; 10^{-6} M_\odot {\rm yr}^{-1}$; therefore, $\dot{M}_{\rm KH}/\dot{M}_{\rm Edd} \approx 10^2$ for the situation of interest. Since mass transfer starts from the more massive component, the actual mass transfer rates are likely to peak at even higher rates. Assuming the binary follows this Cygnus X–2 like evolution, and that the ejected material carries with it the specific angular momentum of the white dwarf, we have the following expression for the relative change in separation due to mass transfer from the Roche-lobe-filling helium star on to the white dwarf

$$F_{\rm CE \to PSN} = \left(\frac{M_{2,\rm CE}}{M_{2,\rm PSN}}\right)^2 \frac{M_{\rm WD} + M_{2,\rm CE}}{M_{\rm WD} + M_{2,\rm PSN}} \times \exp\left[\frac{2(M_{2,\rm PSN} - M_{2,\rm CE})}{M_{\rm WD}}\right] \quad (25)$$

where $M_{2,\rm PSN}$ is evaluated using equation (22). After this phase of mass transfer the core of the helium star explodes as a supernova. Again we synthesize a population of post-supernova binaries which is shown in Fig. 6. It is clear from this figure that J1141–6545–like systems can be produced via this route but that B2023+46–like systems cannot be a product of systems which have undergone mass transfer after the common envelope phase.

We therefore see two distinct evolutionary paths: systems undergoing mass transfer after a common envelope phase produce systems like J1141–6545 whilst those like B2023+46 are produced when the pre-supernova helium star fails to fill its Roche lobe. All computations of He star evolution agree on the fact that for single He stars $R_{\rm max,He}(M)$ drops *rapidly* from $\approx 10^2 \; R_\odot$ to a few $R_\odot$ between $M_{\rm He} \approx 2.5 \; M_\odot$ and $\approx 4 M_\odot$. It is this rapid drop in $R_{\rm max,He}(M)$ which leads to the clear division between the evolutionary channels for J1141–type and B2303–type systems. Where exactly this dividing line is is less important than that it exists.

## 5 DISCUSSION

### 5.1 System birthrates

We now consider the production of J1141–6545–like and B2023+46–like objects from binaries containing a range of primary masses. As can be seen from Fig. 1, constraints on the initial masses of the two stars (as decribed in section 2.1) restrict the primary mass to $4.6 M_\odot \lesssim M_1 \lesssim 8.3 M_\odot$. In Fig. 7, we show the computed fraction of binaries producing J1141–6545–like and B2023+46–like objects, assuming the binaries to be distributed uniformly in $\log(a_{\rm i}/R_\odot)$ and $1 \leq \log(a_{\rm i}) \leq 3$. From this figure we note that the further restrictions on the parameter space produce the two types of binary (as described in section 2.2) are such that no binaries of either type are produced for $M_1 \lesssim 6.4 M_\odot$. This in turn implies that the mass of white dwarfs generated in this way, $M_{\rm WD} \gtrsim 1 \; M_\odot$. We note also that the fraction of binaries producing J1141–6545–like systems is far higher than that for B2023+46–like systems. As also shown in Fig. 7, we considered different distributions of secondary masses, ie different values of $\alpha$ where $dN_2/dM_2 \propto M_2^{-\alpha}$. A flat distribution (ie $\alpha = 0$) produced a larger fraction of both types of binary compared to a Salpeter–like distribution (ie $\alpha = 2.35$). This is because larger secondary masses are required to produce either J1141–6545–like or B2023+46–like systems.

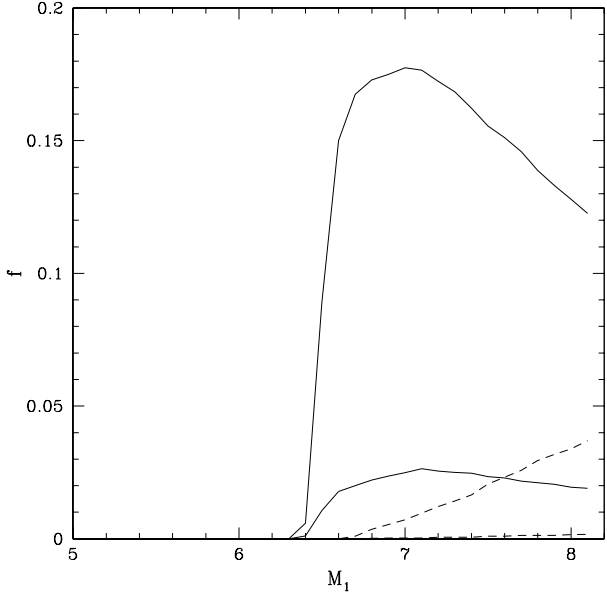

**Figure 7.** The fraction of binary systems, $f$, producing either an J1141–6545–like system (solid lines) or a B2023+46–like system (dashed lines). In each case the lower of the two lines assumes a Salpeter–like distribution of secondary masses, whilst the upper line assumes a flat distribution of secondary masses.

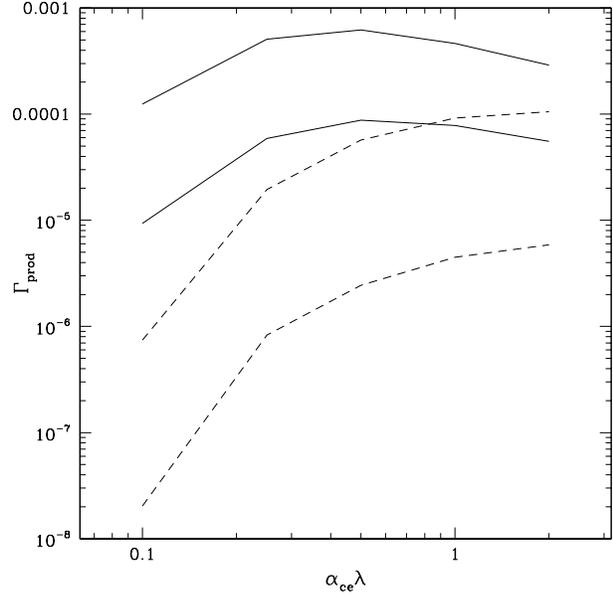

**Figure 8.** The galactic production rate (in units of systems $\mathrm{yr}^{-1}$) of J1141–6545–like systems (solid lines) and B2023+46–like systems (dashed lines) as a function of $\alpha_{ce}\lambda$. In each case the lower of the two lines assumes a Salpeter–like distribution of secondary masses, whilst the upper line assumes a flat distribution of secondary masses.

| $\alpha$ | $\dot{N}_{1141}$ | $\dot{N}_{2303}$ |
|---|---|---|
| 2.35  | $8.75 \times 10^{-5}$ | $2.45 \times 10^{-6}$ |
| 0.00  | $6.20 \times 10^{-4}$ | $5.70 \times 10^{-5}$ |
| -2.35 | $7.60 \times 10^{-4}$ | $1.17 \times 10^{-4}$ |

**Table 1.** The galactic production rate (in units of systems $\mathrm{yr}^{-1}$) of J1141–6545–like and B2023+46–like systems ($\dot{N}_{1141}$ and $\dot{N}_{2303}$ respectively) as a function of distribution of secondary masses, $\alpha$, where $dN_2/dM_2 \propto M_2^{-\alpha}$.

We can now compute the galactic formation rate of systems assuming a primary formation rate $d\dot{N}_1 = M_1^{-2.35} dM_1$ (which is equivalent to a galactic star formation rate of $\sim 3 M_\odot/\mathrm{yr}$). We also assume that one quarter of all massive stars are formed in binaries having the separation range $1 \leq \log(a_i) \leq 3$. Hence the formation rate of systems of type X, derived from binaries having an original primary mass $M_1$, is $d\dot{N}_{1,X} = f_X d\dot{N}_1/4$, where $f_X$ is the fraction of binaries, having the separation range $1 \leq \log(a_i) \leq 3$, which produce systems of type X (as shown in Fig. 7). In Table 1 we list the galactic formation rate of systems as a function of $\alpha$, for $\alpha_{ce}\lambda = 0.5$. We note that for both J1141–6545–like and B2023+46–like systems, the formation rate increases with decreasing $\alpha$, as a larger fraction of the original binaries contain more massive secondaries. The relative formation rates of the two types of systems is also a function of $\alpha$ as B2023+46–like systems require a secondary of mass similar to that of the primary (as shown in Fig. 4) whereas J1141–6545–like systems are produced for a broader range of secondary masses. We now consider the effect of changing the value of $\alpha_{ce}\lambda$. In Fig. 8 we plot the galactic production rates for J1141–6545–like and B2023+46–like systems, for the two cases $\alpha = 2.35$ (ie Salpeter–like distribution of secondary masses) and $\alpha = 0.0$ (ie a flat distribution of secondary masses). The production rate of J1141–6545–like systems is relatively independent of $\alpha_{ce}\lambda$ whereas the production rate of B2023+46–like systems increases with $\alpha_{ce}\lambda$. The formation rate of J1141–6545–like systems appears to be larger than for B2023+46–like systems for all reasonable values of $\alpha$ and $\alpha_{ce}\lambda$.

We now consider the formation rates inferred from the two observed systems. J1141–6545 is at a distance $\simeq 3$ kpc and has a characteristic age derived from the pulsar spin-down of 1.4 Myr. The distance to B2023+46 is less well-known but is thought to lie between $3 - 10$ kpc. It has a spin-down age of 30 Myr. The beaming of pulsar emission means that we will only see those which are pointing towards us. We might only see 1/5 of the actual population. Including the effect of beaming, the distances to both systems would suggest that there are currently $50 - 500$ similar systems thoughout the Galaxy. From the pulsar characteristic ages, this in turn implies formation rates of $5 \times 10^{-5} - 5 \times 10^{-4}$ $\mathrm{yr}^{-1}$ for J1141–6545–like systems and $2 \times 10^{-6} - 2 \times 10^{-5}$ $\mathrm{yr}^{-1}$ for B2023+46–like systems. These values are consistent with the range seen in Table 1. From Figure 8, we note that in addition that the observed formation rates of B2023+46–like systems would seem to suggest that $\alpha_{ce}\lambda \gtrsim 0.3$. We therefore conclude that the scenario described in this paper

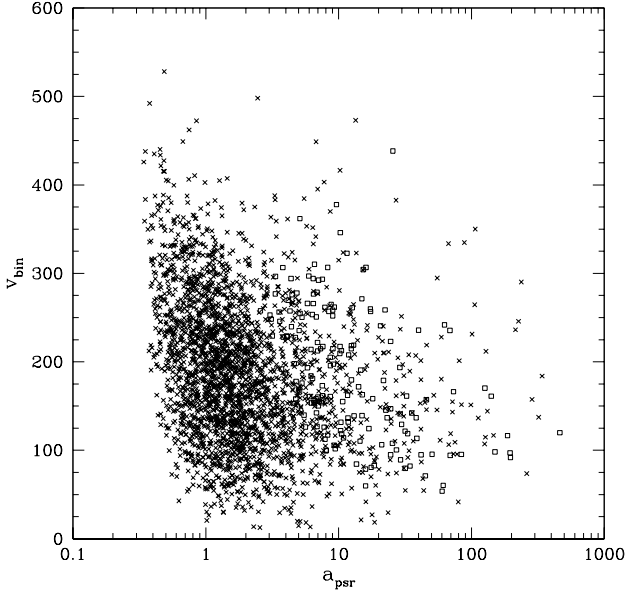

**Figure 9.** The kick velocity received by the WD–NS binaries $V_{\rm bin}$ (in km/s) as a function of their semi-major axes $a_{\rm psr}$ (in solar radii). Crosses are for systems produced via the J1141–6545–like route, whilst the open squares are for systems produced via the B2023+46–like route.

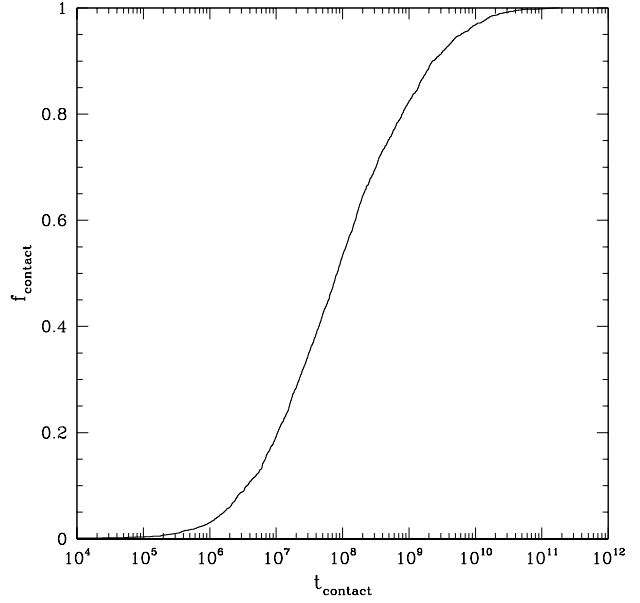

**Figure 10.** The cumulative distribution of timescales for the population of inspiralling WD–NS binaries to come in to contact via the emission of gravitational radiation, $t_{\rm contact}$ (in years).

is able to account for the two observed systems PSR J1141–6545 and B2023+46.

### 5.2 Kick velocities

We consider now the system kick received by J1141–6545–like and B2023+46–like binaries. The system kick of each binary is a combination of the kick the neutron star receives at birth with the kick resulting from the mass–loss in the supernova explosion of the helium star. The distribution of system kick velocities are shown in Fig. 9. From this figure we note that the kicks span a large range of values, the largest being likely to remove the system from the Galaxy entirely (when combined appropriately with the initial orbital velocity of the pre-supernova binary). In a majority of cases, such kicks will at least take the binary significantly away from the plane of the galactic disk (where the systems will have been formed). The kicks for systems most closely resembling J1141–6545 range between 50 km/s and 300 km/s. For B2023+46 the equivalent range is 100 km/s to 200 km/s. These are both consistent with the observed positions of J1141–6545 and B2023+46. J1141–6545 has a galactic latitude of −3.86 degrees which, at a distance of 3.2 kpc and assuming an age of 1.4 Myr, implies a velocity out of the galactic plane $\lesssim 150$ km/s. B2023+46 has a galactic latitude of −12.02 degrees which, at a distance of 3 − 10 kpc and assuming an age of 30 Myr, implies a velocity out of the galactic plane of $\sim 20 - 60$ km/s.

### 5.3 Merger rates

In this section we discuss the subsequent evolution of the J1141–6545–like objects. As can be seen from Fig. 6, in these systems, the white dwarf and neutron star have separations $\sim 1-10 \rm R_\odot$, and a wide range of eccentricities. These systems will spiral in as they lose angular momentum and energy via the emission of gravitational radiation. One may compute the subsequent evolution of the binary separation, $a$, and eccentricity, $e$, using the following expressions (Peters 1964)

$$\frac{da}{dt} = -\frac{64G^3 M_{\rm WD} M_{\rm NS}(M_{\rm WD} + M_{\rm NS})}{5c^5 a^3 (1-e^2)^{7/2}} \times \left(1 + \frac{73}{24}e^2 + \frac{37}{96}e^4\right) \quad (26)$$

$$\frac{de}{dt} = -\frac{304G^3 M_{\rm WD} M_{\rm NS}(M_{\rm WD} + M_{\rm NS})e}{15c^5 a^4 (1-e^2)^{5/2}} \times \left(1 + \frac{121}{304}e^2\right) (27)$$

We thus computed the time required for the binaries to evolve into contact. The cumulative distribution for contact times is shown in Fig. 10. We see from this figure that over half of the systems will merge in a timecale $\lesssim 10^8$ years, and more than 95% within a Hubble time. These merging systems may make an important contribution to the population of gamma-ray burst progenitors as will be explored in a later paper.

## 6 SUMMARY

We have described an evolutionary pathway for producing white dwarf–neutron star binaries where the white dwarf is made first. This explains the observed systems J1141–6545 and B2023+46. In this scenario, the original primary transfers its envelope to the secondary conservatively during a radiative Case B phase of mass transfer leaving only the helium-star core. This helium star then evolves, filling its Roche lobe and leading to the second phase of mass transfer (Case BB) where the envelope of the helium star is transferred to the secondary (which is still on the main sequence). In the process the primary becomes a CO or ONeMg white dwarf. The secondary evolves into contact on its nuclear evolutionary timescale. If the system is sufficiently wide, this third phase of mass transfer leads to a common-envelope phase during which the white dwarf and the helium-star core of the secondary spiral together as the common envelope of gas is ejected from the system. The post-common-envelope system consists of the white dwarf and the helium star in a tight binary. Providing sufficient mass has been transferred to the secondary during earlier phases in the binary evolution, the helium star may be massive enough to produce a neutron star via a supernova explosion. The post-common-envelope but pre-supernova evolution depends on the separation of the binary.

Systems like B2023+46 are produced in relatively wide systems, where the helium star avoids filling its Roche lobe as it expands. Rather it loses mass via a wind. The core of the helium star eventually explodes as a supernova producing a neutron star. Assuming the neutron star receives a kick at birth, the majority of these systems are broken up, producing high-velocity, single, white dwarfs and neutron stars. In a minority of cases, the neutron star and white dwarf produce a binary having a separation $\sim 5 - 50 R_\odot$.

For closer binaries, the helium star in the post-common-envelope system fills its Roche lobe. In this paper, we have proposed that a phase of mass transfer follows where the helium star envelope is transferred to the white dwarf at highly super-Eddington rates. Rather than being accreted, we propose that this material is ejected from the system. Subsequently the helium-star core explodes as a supernova, producing a neutron star. As these binaries are significantly tighter than those producing B2023+46–like systems, the vast majority remain bound, having separations in the range $\sim 0.5 - 5 R_\odot$. This is the way we believe J1141–6545 was formed.

We have computed the expected formation rates for J1141–6545–like and B2023+46–like systems and shown that they are both large and consistent with the two observed systems.

We have shown that the vast majority of J1141–6545–like binaries will spiral in and merge in less than a Hubble time due to the emission of gravitational radiation. Given their relatively high formation rate ($\sim 10^{-4} - 10^{-5}$ yr$^{-1}$), these systems may make an important contribution to the population of gamma-ray burst progenitors.

**Acknowledgments** MBD gratefully acknowledges the support of a URF from the Royal Society. Theoretical astrophysics research at Leicester is supported by a rolling grant from the UK Particle Physics & Astronomy Research Council.